\def\PsfigVersion{1.10}
\def\setDriver{\DvipsDriver} 
\ifx\undefined\psfig\else\endinput\fi
\let\LaTeXAtSign=\@
\let\@=\relax
\edef\psfigRestoreAt{\catcode`\@=\number\catcode`@\relax}
\catcode`\@=11\relax
\newwrite\@unused
\def\ps@typeout#1{{\let\protect\string\immediate\write\@unused{#1}}}

\def\DvipsDriver{
	\ps@typeout{psfig/tex \PsfigVersion -dvips}
\def\PsfigSpecials{\DvipsSpecials} 	\def\ps@dir{/}
\def\ps@predir{} }
\def\OzTeXDriver{
	\ps@typeout{psfig/tex \PsfigVersion -oztex}
	\def\PsfigSpecials{\OzTeXSpecials}
	\def\ps@dir{:}
	\def\ps@predir{:}
	\catcode`\^^J=5
}

\def\figurepath{./:}

\def\DoPaths#1{\expandafter\EachPath#1\stoplist}
\def\leer{}
\def\EachPath#1:#2\stoplist{
  \ExistsFile{#1}{\SearchedFile}
  \ifx#2\leer
  \else
    \expandafter\EachPath#2\stoplist
  \fi}
\def\ps@dir{/}
\def\ExistsFile#1#2{%
   \openin1=\ps@predir#1\ps@dir#2
   \ifeof1
       \closein1
   \else
       \closein1
        \ifx\ps@founddir\leer
           \edef\ps@founddir{#1}
        \fi
   \fi}
\def\get@dir#1{%
  \def\ps@founddir{}
  \def\SearchedFile{#1}
  \DoPaths\figurepath
}

\def\@nnil{\@nil}
\def\@empty{}
\def\@psdonoop#1\@@#2#3{}
\def\@psdo#1:=#2\do#3{\edef\@psdotmp{#2}\ifx\@psdotmp\@empty \else
    \expandafter\@psdoloop#2,\@nil,\@nil\@@#1{#3}\fi}
\def\@psdoloop#1,#2,#3\@@#4#5{\def#4{#1}\ifx #4\@nnil \else
       #5\def#4{#2}\ifx #4\@nnil \else#5\@ipsdoloop #3\@@#4{#5}\fi\fi}
\def\@ipsdoloop#1,#2\@@#3#4{\def#3{#1}\ifx #3\@nnil 
       \let\@nextwhile=\@psdonoop \else
      #4\relax\let\@nextwhile=\@ipsdoloop\fi\@nextwhile#2\@@#3{#4}}
\def\@tpsdo#1:=#2\do#3{\xdef\@psdotmp{#2}\ifx\@psdotmp\@empty \else
    \@tpsdoloop#2\@nil\@nil\@@#1{#3}\fi}
\def\@tpsdoloop#1#2\@@#3#4{\def#3{#1}\ifx #3\@nnil 
       \let\@nextwhile=\@psdonoop \else
      #4\relax\let\@nextwhile=\@tpsdoloop\fi\@nextwhile#2\@@#3{#4}}
\ifx\undefined\fbox
\newdimen\fboxrule
\newdimen\fboxsep
\newdimen\ps@tempdima
\newbox\ps@tempboxa
\long\def\fbox#1{\leavevmode\setbox\ps@tempboxa\hbox{#1}\ps@tempdima\fboxrule
    \advance\ps@tempdima \fboxsep \advance\ps@tempdima \dp\ps@tempboxa
   \hbox{\lower \ps@tempdima\hbox
  {\vbox{\hrule height \fboxrule
          \hbox{\vrule width \fboxrule \hskip\fboxsep
          \vbox{\vskip\fboxsep \box\ps@tempboxa\vskip\fboxsep}\hskip 
                 \fboxsep\vrule width \fboxrule}
                 \hrule height \fboxrule}}}}
\fi
\newread\ps@stream
\newif\ifnot@eof       
\newif\if@noisy        
\newif\if@atend        
\newif\if@psfile       
{\catcode`\%=12\global\gdef\epsf@start{
\def\epsf@PS{PS}
\def\epsf@getbb#1{%
\openin\ps@stream=\ps@predir#1
\ifeof\ps@stream\ps@typeout{Error, File #1 not found}\else
   {\not@eoftrue \chardef\other=12
    \def\do##1{\catcode`##1=\other}\dospecials \catcode`\ =10
    \loop
       \if@psfile
	  \read\ps@stream to \epsf@fileline
       \else{
	  \obeyspaces
          \read\ps@stream to \epsf@tmp\global\let\epsf@fileline\epsf@tmp}
       \fi
       \ifeof\ps@stream\not@eoffalse\else
       \if@psfile\else
       \expandafter\epsf@test\epsf@fileline:. \\%
       \fi
          \expandafter\epsf@aux\epsf@fileline:. \\%
       \fi
   \ifnot@eof\repeat
   }\closein\ps@stream\fi}%
\long\def\epsf@test#1#2#3:#4\\{\def\epsf@testit{#1#2}
			\ifx\epsf@testit\epsf@start\else
\ps@typeout{Warning! File does not start with `\epsf@start'.  It may not be a PostScript file.}
			\fi
			\@psfiletrue} 
{\catcode`\%=12\global\let\epsf@percent=
\long\def\epsf@aux#1#2:#3\\{\ifx#1\epsf@percent
   \def\epsf@testit{#2}\ifx\epsf@testit\epsf@bblit
	\@atendfalse
        \epsf@atend #3 . \\%
	\if@atend	
	   \if@verbose{
		\ps@typeout{psfig: found `(atend)'; continuing search}
	   }\fi
        \else
        \epsf@grab #3 . . . \\%
        \not@eoffalse
        \global\no@bbfalse
        \fi
   \fi\fi}%
\def\epsf@grab #1 #2 #3 #4 #5\\{%
   \global\def\epsf@llx{#1}\ifx\epsf@llx\empty
      \epsf@grab #2 #3 #4 #5 .\\\else
   \global\def\epsf@lly{#2}%
   \global\def\epsf@urx{#3}\global\def\epsf@ury{#4}\fi}%
\def\epsf@atendlit{(atend)} 
\def\epsf@atend #1 #2 #3\\{%
   \def\epsf@tmp{#1}\ifx\epsf@tmp\empty
      \epsf@atend #2 #3 .\\\else
   \ifx\epsf@tmp\epsf@atendlit\@atendtrue\fi\fi}

\chardef\psletter = 11 
\chardef\other = 12

\newif \ifdebug 
\newif\ifc@mpute 
\c@mputetrue 

\let\then = \relax
\def\r@dian{pt }
\let\r@dians = \r@dian
\let\dimensionless@nit = \r@dian
\let\dimensionless@nits = \dimensionless@nit
\def\internal@nit{sp }
\let\internal@nits = \internal@nit
\newif\ifstillc@nverging
\def \Mess@ge #1{\ifdebug \then \message {#1} \fi}

{ 
	\catcode `\@ = \psletter
	\gdef \nodimen {\expandafter \n@dimen \the \dimen}
	\gdef \term #1 #2 #3%
	       {\edef \t@ {\the #1}
		\edef \t@@ {\expandafter \n@dimen \the #2\r@dian}%
		\t@rm {\t@} {\t@@} {#3}%
	       }
	\gdef \t@rm #1 #2 #3%
	       {{%
		\count 0 = 0
		\dimen 0 = 1 \dimensionless@nit
		\dimen 2 = #2\relax
		\Mess@ge {Calculating term #1 of \nodimen 2}%
		\loop
		\ifnum	\count 0 < #1
		\then	\advance \count 0 by 1
			\Mess@ge {Iteration \the \count 0 \space}%
			\Multiply \dimen 0 by {\dimen 2}%
			\Mess@ge {After multiplication, term = \nodimen 0}%
			\Divide \dimen 0 by {\count 0}%
			\Mess@ge {After division, term = \nodimen 0}%
		\repeat
		\Mess@ge {Final value for term #1 of 
				\nodimen 2 \space is \nodimen 0}%
		\xdef \Term {#3 = \nodimen 0 \r@dians}%
		\aftergroup \Term
	       }}
	\catcode `\p = \other
	\catcode `\t = \other
	\gdef \n@dimen #1pt{#1} 
}

\def \Divide #1by #2{\divide #1 by #2} 

\def \Multiply #1by #2
       {{
	\count 0 = #1\relax
	\count 2 = #2\relax
	\count 4 = 65536
	\Mess@ge {Before scaling, count 0 = \the \count 0 \space and
			count 2 = \the \count 2}%
	\ifnum	\count 0 > 32767 
	\then	\divide \count 0 by 4
		\divide \count 4 by 4
	\else	\ifnum	\count 0 < -32767
		\then	\divide \count 0 by 4
			\divide \count 4 by 4
		\else
		\fi
	\fi
	\ifnum	\count 2 > 32767 
	\then	\divide \count 2 by 4
		\divide \count 4 by 4
	\else	\ifnum	\count 2 < -32767
		\then	\divide \count 2 by 4
			\divide \count 4 by 4
		\else
		\fi
	\fi
	\multiply \count 0 by \count 2
	\divide \count 0 by \count 4
	\xdef \product {#1 = \the \count 0 \internal@nits}%
	\aftergroup \product
       }}

\def\r@duce{\ifdim\dimen0 > 90\r@dian \then   
		\multiply\dimen0 by -1
		\advance\dimen0 by 180\r@dian
		\r@duce
	    \else \ifdim\dimen0 < -90\r@dian \then  
		\advance\dimen0 by 360\r@dian
		\r@duce
		\fi
	    \fi}

\def\Sine#1%
       {{%
	\dimen 0 = #1 \r@dian
	\r@duce
	\ifdim\dimen0 = -90\r@dian \then
	   \dimen4 = -1\r@dian
	   \c@mputefalse
	\fi
	\ifdim\dimen0 = 90\r@dian \then
	   \dimen4 = 1\r@dian
	   \c@mputefalse
	\fi
	\ifdim\dimen0 = 0\r@dian \then
	   \dimen4 = 0\r@dian
	   \c@mputefalse
	\fi
	\ifc@mpute \then
		\divide\dimen0 by 180
		\dimen0=3.141592654\dimen0
		\dimen 2 = 3.1415926535897963\r@dian 
		\divide\dimen 2 by 2 
		\Mess@ge {Sin: calculating Sin of \nodimen 0}%
		\count 0 = 1 
		\dimen 2 = 1 \r@dian 
		\dimen 4 = 0 \r@dian 
		\loop
			\ifnum	\dimen 2 = 0 
			\then	\stillc@nvergingfalse 
			\else	\stillc@nvergingtrue
			\fi
			\ifstillc@nverging 
			\then	\term {\count 0} {\dimen 0} {\dimen 2}%
				\advance \count 0 by 2
				\count 2 = \count 0
				\divide \count 2 by 2
				\ifodd	\count 2 
				\then	\advance \dimen 4 by \dimen 2
				\else	\advance \dimen 4 by -\dimen 2
				\fi
		\repeat
	\fi		
			\xdef \sine {\nodimen 4}%
       }}

\def\Cosine#1{\ifx\sine\UnDefined\edef\Savesine{\relax}\else
		             \edef\Savesine{\sine}\fi
	{\dimen0=#1\r@dian\advance\dimen0 by 90\r@dian
	 \Sine{\nodimen 0}
	 \xdef\cosine{\sine}
	 \xdef\sine{\Savesine}}}	      

\def\psdraft{
	\def\@psdraft{0}
}
\def\psfull{
	\def\@psdraft{100}
}

\psfull

\newif\if@scalefirst
\def\psscalefirst{\@scalefirsttrue}
\def\psrotatefirst{\@scalefirstfalse}
\psrotatefirst

\newif\if@draftbox
\def\psnodraftbox{
	\@draftboxfalse
}
\def\psdraftbox{
	\@draftboxtrue
}
\@draftboxtrue

\newif\if@prologfile
\newif\if@postlogfile
\def\pssilent{
	\@noisyfalse
}
\def\psnoisy{
	\@noisytrue
}
\psnoisy
\newif\if@bbllx
\newif\if@bblly
\newif\if@bburx
\newif\if@bbury
\newif\if@height
\newif\if@width
\newif\if@rheight
\newif\if@rwidth
\newif\if@angle
\newif\if@clip
\newif\if@verbose
\def\@p@@sclip#1{\@cliptrue}
\newif\if@decmpr
\def\@p@@sfigure#1{\def\@p@sfile{null}\def\@p@sbbfile{null}\@decmprfalse
   \openin1=\ps@predir#1
   \ifeof1
	\closein1
	\get@dir{#1}
	\ifx\ps@founddir\leer
		\openin1=\ps@predir#1.bb
		\ifeof1
			\closein1
			\get@dir{#1.bb}
			\ifx\ps@founddir\leer
				\ps@typeout{Can't find #1 in \figurepath}
			\else
				\@decmprtrue
				\def\@p@sfile{\ps@founddir\ps@dir#1}
				\def\@p@sbbfile{\ps@founddir\ps@dir#1.bb}
			\fi
		\else
			\closein1
			\@decmprtrue
			\def\@p@sfile{#1}
			\def\@p@sbbfile{#1.bb}
		\fi
	\else
		\def\@p@sfile{\ps@founddir\ps@dir#1}
		\def\@p@sbbfile{\ps@founddir\ps@dir#1}
	\fi
   \else
	\closein1
	\def\@p@sfile{#1}
	\def\@p@sbbfile{#1}
   \fi
}
\def\@p@@sfile#1{\@p@@sfigure{#1}}
\def\@p@@sbbllx#1{
		\@bbllxtrue
		\dimen100=#1
		\edef\@p@sbbllx{\number\dimen100}
}
\def\@p@@sbblly#1{
		\@bbllytrue
		\dimen100=#1
		\edef\@p@sbblly{\number\dimen100}
}
\def\@p@@sbburx#1{
		\@bburxtrue
		\dimen100=#1
		\edef\@p@sbburx{\number\dimen100}
}
\def\@p@@sbbury#1{
		\@bburytrue
		\dimen100=#1
		\edef\@p@sbbury{\number\dimen100}
}
\def\@p@@sheight#1{
		\@heighttrue
		\dimen100=#1
   		\edef\@p@sheight{\number\dimen100}
}
\def\@p@@swidth#1{
		\@widthtrue
		\dimen100=#1
		\edef\@p@swidth{\number\dimen100}
}
\def\@p@@srheight#1{
		\@rheighttrue
		\dimen100=#1
		\edef\@p@srheight{\number\dimen100}
}
\def\@p@@srwidth#1{
		\@rwidthtrue
		\dimen100=#1
		\edef\@p@srwidth{\number\dimen100}
}
\def\@p@@sangle#1{
		\@angletrue
		\edef\@p@sangle{#1} 
}
\def\@p@@ssilent#1{ 
		\@verbosefalse
}
\def\@p@@sprolog#1{\@prologfiletrue\def\@prologfileval{#1}}
\def\@p@@spostlog#1{\@postlogfiletrue\def\@postlogfileval{#1}}
\def\@cs@name#1{\csname #1\endcsname}
\def\@setparms#1=#2,{\@cs@name{@p@@s#1}{#2}}
\def\ps@init@parms{
		\@bbllxfalse \@bbllyfalse
		\@bburxfalse \@bburyfalse
		\@heightfalse \@widthfalse
		\@rheightfalse \@rwidthfalse
		\def\@p@sbbllx{}\def\@p@sbblly{}
		\def\@p@sbburx{}\def\@p@sbbury{}
		\def\@p@sheight{}\def\@p@swidth{}
		\def\@p@srheight{}\def\@p@srwidth{}
		\def\@p@sangle{0}
		\def\@p@sfile{} \def\@p@sbbfile{}
		\def\@p@scost{10}
		\def\@sc{}
		\@prologfilefalse
		\@postlogfilefalse
		\@clipfalse
		\if@noisy
			\@verbosetrue
		\else
			\@verbosefalse
		\fi
}
\def\parse@ps@parms#1{
	 	\@psdo\@psfiga:=#1\do
		   {\expandafter\@setparms\@psfiga,}}
\newif\ifno@bb
\def\bb@missing{
	\if@verbose{
		\ps@typeout{psfig: searching \@p@sbbfile \space  for bounding box}
	}\fi
	\no@bbtrue
	\epsf@getbb{\@p@sbbfile}
        \ifno@bb \else \bb@cull\epsf@llx\epsf@lly\epsf@urx\epsf@ury\fi
}	
\def\bb@cull#1#2#3#4{
	\dimen100=#1 bp\edef\@p@sbbllx{\number\dimen100}
	\dimen100=#2 bp\edef\@p@sbblly{\number\dimen100}
	\dimen100=#3 bp\edef\@p@sbburx{\number\dimen100}
	\dimen100=#4 bp\edef\@p@sbbury{\number\dimen100}
	\no@bbfalse
}
\newdimen\p@intvaluex
\newdimen\p@intvaluey
\def\rotate@#1#2{{\dimen0=#1 sp\dimen1=#2 sp
		  \global\p@intvaluex=\cosine\dimen0
		  \dimen3=\sine\dimen1
		  \global\advance\p@intvaluex by -\dimen3
		  \global\p@intvaluey=\sine\dimen0
		  \dimen3=\cosine\dimen1
		  \global\advance\p@intvaluey by \dimen3
		  }}
\def\compute@bb{
		\no@bbfalse
		\if@bbllx \else \no@bbtrue \fi
		\if@bblly \else \no@bbtrue \fi
		\if@bburx \else \no@bbtrue \fi
		\if@bbury \else \no@bbtrue \fi
		\ifno@bb \bb@missing \fi
		\ifno@bb \ps@typeout{FATAL ERROR: no bb supplied or found}
			\no-bb-error
		\fi
		\count203=\@p@sbburx
		\count204=\@p@sbbury
		\advance\count203 by -\@p@sbbllx
		\advance\count204 by -\@p@sbblly
		\edef\ps@bbw{\number\count203}
		\edef\ps@bbh{\number\count204}
		\if@angle 
			\Sine{\@p@sangle}\Cosine{\@p@sangle}
	        	{\dimen100=\maxdimen\xdef\r@p@sbbllx{\number\dimen100}
					    \xdef\r@p@sbblly{\number\dimen100}
			                    \xdef\r@p@sbburx{-\number\dimen100}
					    \xdef\r@p@sbbury{-\number\dimen100}}
                        \def\minmaxtest{
			   \ifnum\number\p@intvaluex<\r@p@sbbllx
			      \xdef\r@p@sbbllx{\number\p@intvaluex}\fi
			   \ifnum\number\p@intvaluex>\r@p@sbburx
			      \xdef\r@p@sbburx{\number\p@intvaluex}\fi
			   \ifnum\number\p@intvaluey<\r@p@sbblly
			      \xdef\r@p@sbblly{\number\p@intvaluey}\fi
			   \ifnum\number\p@intvaluey>\r@p@sbbury
			      \xdef\r@p@sbbury{\number\p@intvaluey}\fi
			   }
			\rotate@{\@p@sbbllx}{\@p@sbblly}
			\minmaxtest
			\rotate@{\@p@sbbllx}{\@p@sbbury}
			\minmaxtest
			\rotate@{\@p@sbburx}{\@p@sbblly}
			\minmaxtest
			\rotate@{\@p@sbburx}{\@p@sbbury}
			\minmaxtest
			\edef\@p@sbbllx{\r@p@sbbllx}\edef\@p@sbblly{\r@p@sbblly}
			\edef\@p@sbburx{\r@p@sbburx}\edef\@p@sbbury{\r@p@sbbury}
		\fi
		\count203=\@p@sbburx
		\count204=\@p@sbbury
		\advance\count203 by -\@p@sbbllx
		\advance\count204 by -\@p@sbblly
		\edef\@bbw{\number\count203}
		\edef\@bbh{\number\count204}
}
\def\in@hundreds#1#2#3{\count240=#2 \count241=#3
		     \count100=\count240	
		     \divide\count100 by \count241
		     \count101=\count100
		     \multiply\count101 by \count241
		     \advance\count240 by -\count101
		     \multiply\count240 by 10
		     \count101=\count240	
		     \divide\count101 by \count241
		     \count102=\count101
		     \multiply\count102 by \count241
		     \advance\count240 by -\count102
		     \multiply\count240 by 10
		     \count102=\count240	
		     \divide\count102 by \count241
		     \count200=#1\count205=0
		     \count201=\count200
			\multiply\count201 by \count100
		 	\advance\count205 by \count201
		     \count201=\count200
			\divide\count201 by 10
			\multiply\count201 by \count101
			\advance\count205 by \count201
		     \count201=\count200
			\divide\count201 by 100
			\multiply\count201 by \count102
			\advance\count205 by \count201
		     \edef\@result{\number\count205}
}
\def\compute@wfromh{
		\in@hundreds{\@p@sheight}{\@bbw}{\@bbh}
		\edef\@p@swidth{\@result}
}
\def\compute@hfromw{
	        \in@hundreds{\@p@swidth}{\@bbh}{\@bbw}
		\edef\@p@sheight{\@result}
}
\def\compute@handw{
		\if@height 
			\if@width
			\else
				\compute@wfromh
			\fi
		\else 
			\if@width
				\compute@hfromw
			\else
				\edef\@p@sheight{\@bbh}
				\edef\@p@swidth{\@bbw}
			\fi
		\fi
}
\def\compute@resv{
		\if@rheight \else \edef\@p@srheight{\@p@sheight} \fi
		\if@rwidth \else \edef\@p@srwidth{\@p@swidth} \fi
}
\def\compute@sizes{
	\compute@bb
	\if@scalefirst\if@angle
	\if@width
	   \in@hundreds{\@p@swidth}{\@bbw}{\ps@bbw}
	   \edef\@p@swidth{\@result}
	\fi
	\if@height
	   \in@hundreds{\@p@sheight}{\@bbh}{\ps@bbh}
	   \edef\@p@sheight{\@result}
	\fi
	\fi\fi
	\compute@handw
	\compute@resv}
\def\OzTeXSpecials{
	\special{empty.ps /@isp {true} def}
	\special{empty.ps \@p@swidth \space \@p@sheight \space
			\@p@sbbllx \space \@p@sbblly \space
			\@p@sbburx \space \@p@sbbury \space
			startTexFig \space }
	\if@clip{
		\if@verbose{
			\ps@typeout{(clip)}
		}\fi
		\special{empty.ps doclip \space }
	}\fi
	\if@angle{
		\if@verbose{
			\ps@typeout{(rotate)}
		}\fi
		\special {empty.ps \@p@sangle \space rotate \space} 
	}\fi
	\if@prologfile
	    \special{\@prologfileval \space } \fi
	\if@decmpr{
		\if@verbose{
			\ps@typeout{psfig: Compression not available
			in OzTeX version \space }
		}\fi
	}\else{
		\if@verbose{
			\ps@typeout{psfig: including \@p@sfile \space }
		}\fi
		\special{epsf=\ps@predir\@p@sfile \space }
	}\fi
	\if@postlogfile
	    \special{\@postlogfileval \space } \fi
	\special{empty.ps /@isp {false} def}
}
\def\DvipsSpecials{
	\special{ps::[begin] 	\@p@swidth \space \@p@sheight \space
			\@p@sbbllx \space \@p@sbblly \space
			\@p@sbburx \space \@p@sbbury \space
			startTexFig \space }
	\if@clip{
		\if@verbose{
			\ps@typeout{(clip)}
		}\fi
		\special{ps:: doclip \space }
	}\fi
	\if@angle
		\if@verbose{
			\ps@typeout{(clip)}
		}\fi
		\special {ps:: \@p@sangle \space rotate \space} 
	\fi
	\if@prologfile
	    \special{ps: plotfile \@prologfileval \space } \fi
	\if@decmpr{
		\if@verbose{
			\ps@typeout{psfig: including \@p@sfile.Z \space }
		}\fi
		\special{ps: plotfile "`zcat \@p@sfile.Z" \space }
	}\else{
		\if@verbose{
			\ps@typeout{psfig: including \@p@sfile \space }
		}\fi
		\special{ps: plotfile \@p@sfile \space }
	}\fi
	\if@postlogfile
	    \special{ps: plotfile \@postlogfileval \space } \fi
	\special{ps::[end] endTexFig \space }
}
\def\psfig#1{\vbox {
	\ps@init@parms
	\parse@ps@parms{#1}
	\compute@sizes
	\ifnum\@p@scost<\@psdraft{
		\PsfigSpecials 
		\vbox to \@p@srheight sp{
			\hbox to \@p@srwidth sp{
				\hss
			}
		\vss
		}
	}\else{
		\if@draftbox{		
			\hbox{\fbox{\vbox to \@p@srheight sp{
			\vss
			\hbox to \@p@srwidth sp{ \hss 
			 \hss }
			\vss
			}}}
		}\else{
			\vbox to \@p@srheight sp{
			\vss
			\hbox to \@p@srwidth sp{\hss}
			\vss
			}
		}\fi

	}\fi
}}
\psfigRestoreAt
\setDriver
\let\@=\LaTeXAtSign

 \documentstyle[aaspptwo]{article}

\def\ms{$M_\odot$}
\def\e#1{$\times$ 10$^{#1}$}
\def\etal{et al. }
\def\ni{$^{56}$Ni}

\begin{document}

\title{Maximum Brightness and Post-Maximum Decline of Light Curves of SN~Ia:
A Comparison of Theory  and Observations}

\author{Peter H\"oflich}
\affil{Center for Astrophysics, Harvard University, Cambridge, MA 02138, USA}
 \affil{Department of Astronomy, University of Texas, Austin, TX 78712, USA}
 \affil{Institute for Theoretical Physics, U Basel, CH-4056 Basel, Switzerland}

\author{Alexei Khokhlov}
\affil{Laboratory for Computational Physics \& Fluid Dynamics, Cole 6402, NRL}
\affil{Washington DC 20375}

\author{J. Craig Wheeler}
\affil{Department of Astronomy, University of Texas, Austin, TX78712, USA}

\author{Mario Hamuy}
\affil{CTIO, Casilla 603, La Serena, Chile }

\author{Mark M. Phillips}
\affil{CTIO, Casilla 603, La Serena, Chile }

   \and

\author{Nicolas B. Suntzeff}
\affil{CTIO, 950 N. Cherry Ave., Tuscon, AZ 85719, USA }

\begin{abstract}
 We compare the observed correlations between the maximum brightness,
postmaximum
decline rate and color at maximum light of Type Ia supernovae (SN~Ia)
with model predictions.

 The observations
are based on a total of 40 SN~Ia with
 29 SN of the Calan Tololo Supernova Search  and 11 local SN
 which cover  a range
of $\approx 2^m$ in the absolute visual brightness.
 The observed correlations are not  tight, one dimensional relations.
Supernovae with the same postmaximum decline or the same color have a spread in
visual magnitude of $\approx  0.7^m$. The dispersion in the color-magnitude
relation
 may result from uncertainties in the distance determinations or the
interstellar reddening within the host galaxy. The dispersion in the decline
rate-magnitude
relation suggests that an intrinsic spread in the supernova properties exists
 that cannot be accounted for by any single
relation between visual brightness and postmaximum decline.

Theoretical correlations are derived from a grid of
 models which encompasses  delayed detonations,
pulsating delayed detonations, the merging scenario and helium detonations.
 We find that the  observed correlations can be understood in terms
of  explosions of Chandrasekhar mass white dwarfs.
 Our models show  an intrinsic spread in the relations
 of about $0.5^m$ in the maximum brightness and  $\approx 0.1^m$ in the B-V
color.
 Our study provides strong evidence against the mechanism of
 helium detonation for subluminous, red SN~Ia.
\end{abstract}

\section{INTRODUCTION}
 Supernovae of Type Ia are the most luminous stellar  objects and,
in principle, can be used to determine extragalactic distances and the
cosmological parameters.
 Their use as standard candles is based on the assumption that they form a
homogeneous
group. Type Ia Supernovae (SN~Ia) were, however, long suspected not to be
perfectly
homogeneous both from the light curves and the spectra (Pskovskii 1970, 1977,
Barbon, Ciatti \& Rosino  1973, 1990, Branch 1981,  Elias \etal 1985, Frogel
\etal 1987,
Phillips \etal 1987, Cristiani \etal 1992).
 The discovery of the strongly subluminous supernova SN 1991bg established the
existence of a wide range of luminosities among SN~Ia
(Filippenko \etal 1992, Leibundgut \etal 1993). New, uniform data sets of
high quality confirm this diversity (Hamuy \etal\ 1993, Maza \etal\ 1994,
Suntzeff, 1995,
Hamuy et al. 1996).
 From these data, the existence of
 a correlation between the maximum brightness and the shape of
the light curves was established and used to correct for the variations in the
absolute
brightness and to determine $H_o$
 (Phillips 1993, Hamuy \etal 1995,
Riess, Kirshner \& Press 1995).

It is widely accepted that SN~Ia are thermonuclear
explosions of carbon-oxygen white dwarfs (Hoyle \& Fowler 1960).
 The three main scenarios are the explosion 1)  of a Chandrasekhar mass white
dwarf,
(Arnett  1969, Nomoto et al. 1976, 1984 , Khokhlov  1991),
2) of merging white dwarfs (Tutukov \& Yungelson 1983, Iben \& Tutukov 1984,
Webbink 1984),
 and 3)  of a low mass
white dwarf triggered by a helium detonation at its surface as suggested by
  Nomoto \etal  (1980) and  Woosley, Taam \& Weaver (1980).
 Within each scenario
different amount, of $^{56}Ni$ can be produced depending on details of the
progenitor
evolution, presupernovae structure and flame propagation. Because Ni is the
main
energy source for the light curve, the brightness of the models must be
expected to
 vary. Detailed modeling of the LCs shows that they differ both
in their brightness and shape, but their physical correlation differs
depending on the
scenario        (H\"oflich, Khokhlov \& M\"uller 1993). Therefore,
 a comparison between theory and observations can be used to discriminate
explosion scenarios.
The theoretical relation can be further used to  determine $H_o$ independent
from
secondary distance indicators needed in purely empirical determination
(M\"uller \& H\"oflich 1994,
H\"oflich \& Khokhlov 1996, and
references therein).

 In this letter, we compare the observed correlations between maximum
brightness,
the post-maximum decline and colors of the visual light curves of SN~Ia with
theory.
 The post-maximum decline is characterized by the parameter $\Delta M_V(t)$
defined as the difference
between the brightness at maximum light and that  $\sl {t}$ days later.
 The comparison is based on 40 well observed supernoave
 and our light curve calculations of a set of 42 models.
The list of supernovae includes the uniform set 29 supernovae obtained with the
Calan Tololo Supernova Search
(SN1990O, 90T, 90Y, 90af, 91S, 91U, 91ag, 92J, 92K, 92P, 92ae, 92ag, 92al,
92aq, 92au, 92bc, 92bg, 92bh,      92bl,
92bo, 92bp, 92br, 92bs, 93B, 93H, 93O, 93ag, 93ah) and 11 nearby supernovae
(SN1937C, 72E, 80N, 81B, 86G, 89B, 90N, 91T, 91bg,
92A, 94D) (see Hamuy et al. 1996).


\begin{figure}
\psfig{figure=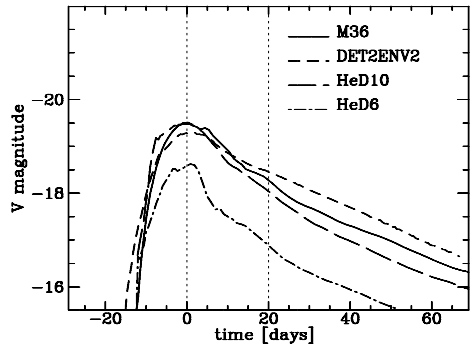,width=6.3cm,rwidth=5.5cm,clip=,angle=270}
 \caption{
         Visual light curves for the delayed detonation model M36, the envelope
models DET2ENV2, and
the helium detonations HeD6 and 10 (from Khokhlov et al. 1993, H\"oflich 1995,
H\"oflich \& Khokhlov 1996).
The two vertical lines mark the time of maximum light and 20 days later. Note,
that for HeD6, $\Delta M_V(20)$
does not provide a good measurement for the post-maximum decline.
}
 \end{figure}

\section{OBSERVATIONS VS. THEORY}

To illustrate the nature of $\Delta M_V(t)$,  we show in figure 1 four
theoretical V light curves
based on different explosion scenarios. The function $\Delta M_V(t)$ provides a
particular measure of
post-maximum decline rate. The color (V)  in which the comparison is made, and
the value of the
time base t must be chosen carefully.
 We use the visual wavelength range because, past
maximum light, most of the energy is emitted in V and, consequently, the
theoretical LCs are most
accurate in V.  Moreover, the spectral variation of the flux across the V
filter is smaller
than in other bands, e.g. B or R. Consequently, differences induced by the
assumed transmission of
filters and those actually used during the observations will be smallest
(H\"oflich 1995).
 We have found that a time base of 15 days, previously used in B by Phillips
(1993), does not permit
a clear distinction
between different visual LCs, because the decline rate in V is much smaller
than in B. Moreover, a value of $\Delta M_V(15)$
is not that sensitive to the postmaximum decline, but is strongly
influenced by the broadness of the maximum. On the other hand,
 a very long base will measure predominantly the exponential decay at late
times.
 We find that t=20 days is
a better choice in order to differentiate the various light curves in V.

 In figure 2, the observed absolute visual brightness $M_V$ is plotted as a
function of
$\Delta M_V(20)$ based on the LCs observed at CTIO (Hamuy et al. 1996).
 The errors are estimated as follows:
 In $M_V$, uncertainties are due to uncertainty in
the apparent magnitude.  For those SN which were
observed at maximum, we estimate an uncertainty of 0.05 mag. For those not
observed
at maximum light, $M_V$ is determined  by  fitting  template curves
to {\sl extrapolate} to a peak magnitude.

\begin{figure}
\psfig{figure=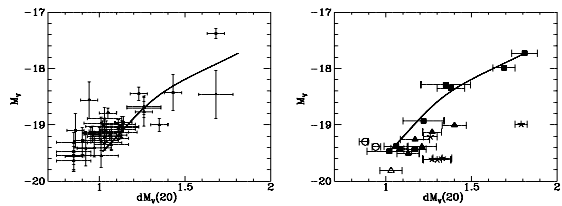,width=6.3cm,rwidth=5.5cm,angle=270}
 \caption{
         Observed $M_V$ as a function $\Delta M_V(20)$ (left plot) normalized
to $H_o = 65 km/(Mpc~s) $
(Hamuy et al. 1995, H\"oflich  \& Khokhlov 1995).
 In the right plot, the theoretical models are shown for the delayed detonation
(open triangle: N-series, Khokhlov 1991; black triangles:
M-series, H\"oflich 1995), pulsating delayed detonations (black circles) and
merging scenarios (open circles)
 (Khokhlov, M\"uller \& H\"oflich 1993,
H\"oflich, Khokhlov \& Wheeler 1995, H\"oflich \& Khokhlov 1995) and the helium
detonations (asteriks,
H\"oflich \& Khokhlov 1996).
 The correlation between $M_V$ and $\Delta M_V(20)$ within each set of models
is evident. Note
that the models of the M- and N-series do produce different relations. Although
both are  based on the delayed detonation mechanism, the flame velocities and
pre-supernova
structures are different.
 The curve represents the theoretical relations for pulsating delayed
detonations given in both plots for orientation.
}
 \end{figure}

 In so doing we are essentially comparing the brightnesses at maximum light
 given by the various templates employed. This
technique provides a way to estimate our uncertainty in guessing a quantity
that was not observed. For instance, if the best fit yields
$V_{max}=15.00^m$ and the next-best-fit yields 15.25, we quote $V_{max}=15.00
\pm 0.25$.
Therefore, our error estimates for the peak luminosities are larger
than 1 $\sigma $ since they cover a range of confidence larger than 67\%.
Given this uncertainty in the peak apparent magnitude, we add in quadrature
an estimate of the foreground extinction correction (0.045 mag), an estimate
in the K-term correction (0.02 mag), and the uncertainty of 600 km s$^{-1}$
in the
 velocity of the cosmological expansion
due to the correction for peculiar motions.
Another source of error, not included in the error bars (see below), is
due to the distance determination of the host galaxies which are based on
Tully-Fisher (1977) and
surface bightness fluctuation (Tonry \& Schneider 1988).
 For $\Delta M_V(20)$, we adopt an error of $0.05 ^m$ for those
SN whose light curves were observed from maximum light through day 20 and
$0.10^m$
for others.

 The correlation between $M_V$ and $\Delta M_V(20)$ can be clearly seen in Fig.
2. With decreasing
brightness at maximum light,
 supernovae decline faster. There is, however, a spread in $M_V$ of about
$0.7^m$
within the relation. This spread is larger than the estimated error. It may be
explained either by
the error in the individual distance determinations, reddening in the host
galaxy,
 or by an intrinsic spread among SN~Ia with
the same $\Delta M_V(20)$, or by a combination of all these effects. In the
first case, this would
imply  an uncertainty of $\approx $ 40 \%  in the distance determinations. This
is much larger
than the relative uncertainties of the Tully-Fisher and the surface brightness
fluctuation which
are 12
\% and 10 \%, respectively (Jacoby et al. 1992).
 The  error in $E_{B-V}$ of the host galaxy of less than $0.1^m$ is probably
realistic.
 Taking the latter error estimates, we are forced to assume an
intrinsic spread of $M_V$ of $\approx 0.3 -- 0.6^m$ of SN~Ia at a given $\Delta
M_V$(20).

 The theoretical relation between  $M_V$ and $\Delta M_V(20)$ is shown on the
right panel of Fig. 2.
 The models do  provide a spread in $M_V$ within each explosion scenario,
and  $\Delta M_V(20)$ decreases with $M_V$. The largest variation is among
delayed  and pulsating delayed
 detonations.  Both scenarios show qualitative agreement with the observations
within the error bars.
For normal bright delayed detonations, however, the post-maximum decline is
somewhat steeper than
observed.
 If this systematic tendency is real, models with a lower central density of
 the exploding white dwarfs may be preferred or expanding envelopes  with a
more pronounced shell-like structure may be favored.
 helium detonations fall well outside the observed range. They
decline much too fast.

 In Figure 3, we give B-V as function of $M_V$ for observed supernovae and our
models.
 With decreasing maximum brightness supernovae become redder (Fig. 3). The
color relation again shows
a substantial scatter. The reasons may be interstellar reddening (Miller \&
Branch 1994),
 errors in the distance determination (see above),
errors intrinsic to the observations,
and/or may reflect an intrinsic spread of properties of Type Ia supernovae.
 Qualitatively, models for the explosion of Chandrasekhar mass white dwarfs
follow, within the
uncertainties, the
same $(B-V)-(M_V)$ relation as the observations. For these models,
 the intrinsic spread of the B-V relation is
apparently of the order of $0.1^m$.  Given the intrinsic uncertainties and
approximations used for the light curve calculations, the discrepancies are
well within the
expected errors. NLTE-effects, for instance, tend to produce slightly bluer
colors
($\approx 0.02 -- 0.05 ^m$) at maximum light compared to our light curve-colors
 (H\"oflich
1995).     Another possible source of systematic errors entering the comparison
is connected to
the filter response functions of the observations and those used for the
theoretical light
curves. For dim supernovae, the models are slightly bluer. This can be
explained by interstellar
reddening, but is more likely due to selective line blanketing (Branch, private
communication)
or dust formation (Dominick et al. 1995, H\"oflich \& Khokhlov 1996).

helium detonations show a rather blue color even if somewhat subluminous. Their
color is
clearly in agreement with bright SN~Ia. A very large reddening would be
required
in order to reproduce the observed extremely subluminous SN~Ia. For SN1992K,
for instance,
$E_{B-V}$ must be as large as $0.7^m$ (Hamuy et al. 1994).
 This would mean an intrinsic brightness of $-20.7^m $
assuming $A_V = 3.1 ~ E_{B-V}$ which is inconsistent with the helium models and
is out  of the
reach of even pure detonation models of Chandrasekhar mass white dwarfs
($M_V=-20^m$, Khokhlov et
al. 1993).

\section{CONCLUSIONS AND DISCUSSION}

     The observed correlations between the absolute brightness and the
postmaximum decline rates
and B-V color can be understood in terms of explosions of Chandrasekhar mass
white dwarfs.
In these models,
the variation in brightness is due to different amounts of $^{56}Ni $ produced
in the
central region. If little Ni is produced, the envelope stays cooler. This has
two effects:
the color is redder and the photosphere recedes faster at maximum light which
results in
a fast postmaximum decline (H\"oflich \& Khokhlov 1996 and references
therein).
\begin{figure}
\psfig{figure=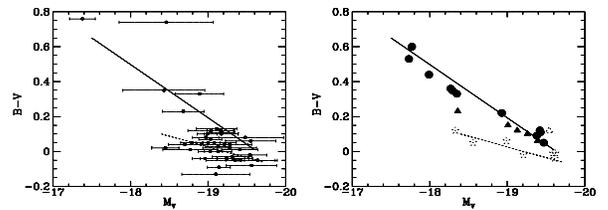,width=6.8cm,rwidth=4.5cm,angle=270}
 \caption{
           Observed (left, normalized to $H_o = 65 km/(Mpc~s) $)
 and theoretical (right) plot of
 B-V  as a function of $M_V$ for the same supernovae and models
 as in Fig. 2.
 The observational errors in the B-V are of the order of a few hundreds of a
magnitude.
 The line show the approximate relation for delayed detonations (solid) of the
M-series
 and helium detonations (dashed).
}
 \end{figure}

 For the very same reason, helium detonations show different behavior.  In
those models, a
significant amount of Ni is present in the outer layers.  This heats up the
photosphere and
keeps it hot even in  subluminous explosions. The color remains blue. This
implies that the red
color observed  in subluminous  SN~Ia must be attributed to interstellar
reddening. This, in
turn, is  incompatible with the maximum brightness (see above).  In addition,
the postmaximum
decline of helium detonations is always steep because,
near maximum light, the outer region with  substantial \ni ~becomes
transparent to $\gamma $ rays. This results in a rapid increase of the escape
probability and,
consequently, in a rapidly declining light curve even for bright SN~Ia.
Note
 that, for normal bright supernovae, early time spectra indicate expansion
velocities
of Si-rich layers in excess of
 19,000 km/sec (e.g. 1990N, Leibundgut et al. 1991; SN1994D, H\"oflich 1995,
SN1995E, Riess,
private communication). In contrast,
both 1-D and 2-D  model calculations for helium detonations predict velocities
smaller than
14,000 km/sec for these layers
(Woosley \& Weaver 1994,  Livne \& Arnett 1995, H\"oflich \& Khokhlov 1996).
 The restriction of Si to low  velocities
 must be regarded as a generic feature of helium detonations.
 Within this scenario, a minimum of
 0.15 to 0.2 $M_\odot$ of He atop the carbon-oxygen
 WD is required.
 Explosive burning of helium
at low densities produces mainly \ni .
 To make helium detonations
consistent with the limits from early spectra both with respect to the
appearance of strong Si lines and
the absence of strong Ni lines,
the burning products of the outer, former He shell
( $M_{He} \approx 0.1~ ...~ 0.2 M_\odot$) must
accelerated  to velocities well above $\approx 16000 ... 18000 km/sec$. The
energy
required would be well in excess of the total energy of a thermonuclear
explosion.
 For more details, see H\"oflich \& Khokhlov (1996).

 Models do not give one-parameter relations for $M_V -\Delta M_V(20)$ and
$(B-V)-M_V$.  If a single
monotonic relation is used (see Figs. 2 \& 3), then a spread exists around this
relation of
0.5$^m$ in $M_V$ and $0.1^m$ in B-V. Thus,  even within a given explosion
scenario, models with
different flame velocities and pre-supernovae structures do  produce the same
$M_V$ but produce
different colors and light curve shapes as a comparison of the models of the N-
and M- series
reveals (Figs. 2 \& 3).

 The observations show an even larger spread in both the $M_V- \Delta M_V(20)$
and $(B-V)-M_V$
relation. This may be partially attributed to uncertainties in the distances
and interstellar
reddening. Within these uncertainties, the observed $(B-V)-M_V$ relation may be
consistent with a
one-parameter relation because the reddening correction enters both B-V and
$M_V$. For
the
$\Delta M_V(20)-M_V$ relation, however, the reddening correction enters $M_V$
only. To attribute the
observed spread in $\Delta M_V(20)-M_V$ to the reddening alone  would require a
mean $E_{B-V}$ of at
least 0.2$^m$.  Based on the statistical studies of Miller \& Branch (1994) and
our individual
fits of SN~Ia light curves, we regard the implied mean reddening as rather
unlikely.
This unacceptably high value indicates that at least a part of the variation is
 intrinsic to SN~Ia. To distangle the different causes of the spread, detailed
analyses of the entire light curves and spectra and deeper understanding of the
physics of the
stellar evolution and explosion is required.

We thank the CTIO supernovae search group for providing data two years
prior to publication.
P.H. would like to thank Bob Kirshner and his group for many helpful
discussions.
A.K. would like to thank  Bob Kirshner for his hospitality at the CfA where
a draft of the paper was written in March 1995.
 This work has been supported by
 grant Ho 1177/2-1 of the Deutsche Forschungsgemeinschaft, by NSF Grant AST
9218035 and by NASA
Grants NAGW 2905 and NAG5-2888.

\end{document}